\documentclass[sigconf,screen]{acmart} 

\usepackage[utf8]{inputenc}        
\usepackage{csquotes}              
\usepackage{enumitem}              
\usepackage[nameinlink]{cleveref}  

\copyrightyear{2022} 
\acmYear{2022} 
\setcopyright{acmlicensed}
\acmConference[SIGCSE 2022]{Proceedings of the 53rd ACM Technical Symposium on Computer Science Education V. 1}{March 3--5, 2022}{Providence, RI, USA}
\acmBooktitle{Proceedings of the 53rd ACM Technical Symposium on Computer Science Education (SIGCSE 2022), March 3--5, 2022, Providence, RI, USA}
\acmPrice{15.00}
\acmDOI{10.1145/3478431.3499414}
\acmISBN{978-1-4503-9070-5/22/03}

\renewenvironment{quote}{\list{}{\leftmargin=1.3em\rightmargin=1.3em}\item\relax\it\hspace{-0.2em}\textquotedblleft\ignorespaces}{\unskip\unskip\textquotedblright\endlist}

\newcommand{\kypo}{\textit{KYPO CRP}}
\newcommand{\edurange}{\textit{EDURange}}
\newcommand{\locust}{\textit{Locust 3302}}
\newcommand{\filewrangler}{\textit{File Wrangler}}
\newcommand{\traineegraph}{\textsc{Trainee graph}}
\newcommand{\milestonegraph}{\textsc{Milestone graph}}

\begin{document}

\fancyhead{} 

\title{Evaluating Two Approaches to Assessing Student Progress in~Cybersecurity Exercises}

\author[V. Švábenský]{Valdemar Švábenský}
\orcid{0000-0001-8546-280X}
\affiliation{
  \institution{Masaryk University}
  \country{Czech Republic}
}
\email{svabensky@ics.muni.cz}

\author[R. Weiss]{Richard Weiss}
\affiliation{
  \institution{The Evergreen State College}
  \country{Washington, USA}
}
\email{weissr@evergreen.edu}

\author[J. Cook]{Jack Cook}
\affiliation{
  \institution{New York University}
  \country{New York, USA}
}
\email{cookjackc@gmail.com}

\author[J. Vykopal]{Jan Vykopal}
\orcid{0000-0002-3425-0951}
\affiliation{
  \institution{Masaryk University}
  \country{Czech Republic}
}
\email{vykopal@ics.muni.cz}

\author[P. Čeleda]{Pavel Čeleda}
\orcid{0000-0002-3338-2856}
\affiliation{
  \institution{Masaryk University}
  \country{Czech Republic}
}
\email{celeda@ics.muni.cz}

\author[J. Mache]{Jens Mache}
\affiliation{
  \institution{Lewis \& Clark College}
  \country{Oregon, USA}
}
\email{jmache@lclark.edu}

\author[R. Chudovský]{Radoslav Chudovský}
\affiliation{
  \institution{Masaryk University}
  \country{Czech Republic}
}
\email{chudovsky@mail.muni.cz}

\author[A. Chattopadhyay]{Ankur Chattopadhyay}
\affiliation{
  \institution{Northern Kentucky University}
  \country{Kentucky, USA}
}
\email{chattopada1@nku.edu}

\begin{abstract}
Cybersecurity students need to develop practical skills such as using command-line tools. Hands-on exercises are the most direct way to assess these skills, but assessing students' mastery is a challenging task for instructors. We aim to alleviate this issue by modeling and visualizing student progress automatically throughout the exercise. The progress is summarized by graph models based on the shell commands students typed to achieve discrete tasks within the exercise. We implemented two types of models and compared them using data from 46 students at two universities. To evaluate our models, we surveyed 22 experienced computing instructors and qualitatively analyzed their responses. The majority of instructors interpreted the graph models effectively and identified strengths, weaknesses, and assessment use cases for each model. Based on the evaluation, we provide recommendations to instructors and explain how our graph models innovate teaching and promote further research. The impact of this paper is threefold. First, it demonstrates how multiple institutions can collaborate to share approaches to modeling student progress in hands-on exercises. Second, our modeling techniques generalize to data from different environments to support student assessment, even outside the cybersecurity domain. Third, we share the acquired data and open-source software so that others can use the models in their classes or research.
\end{abstract}



\maketitle

\section{Introduction}
\label{sec:intro}

Cybersecurity is an essential topic in the ACM/IEEE Computing Curricula 2020~\cite{cc2020}. However, it is challenging for students to learn since it encompasses skills from many areas of computing, such as programming, operating systems, and networking. To promote deep understanding, students must practice these skills hands-on.

Subsequent assessment of students' learning is vital~\cite{handbook-CER14}. However, the assessment of practical tasks is challenging for several reasons. If it is performed manually, it is time-consuming and can be inaccurate due to the quantity and complexity of student interaction data. If automated, it is often superficial, including only the information about whether the solution was correct or not~\cite{weiss2016}.

To help instructors overcome this challenge, we propose and evaluate two methods for supporting semi-automated, timely, accurate, and in-depth assessment of students. The methods are based on visualizing command-line histories from solving cybersecurity tasks, resulting in graphical progress models. Instructors can use these models to better understand how their students learn. For example, they can compare the students' approaches to solving the tasks, along with the mistakes they made. Based on this understanding, the instructors can assess students in two ways~\cite{handbook-CER14}:

\begin{itemize}[leftmargin=5mm]
    \item \textit{formatively} -- providing feedback to students to support their learning, for example, correcting the struggling students, and
    \item \textit{summatively} -- grading students to evaluate their level of knowledge, for example, distinguishing advanced students and novices.
\end{itemize}

This paper follows the multi-national, multi-institutional study framework, which addresses the limitations of many computing education research papers~\cite{handbook-CER1}. We employ two different interactive learning environments, exercises, and student/instructor populations from two continents. Using the methods of educational data mining~\cite{handbook-edm2010} and learning analytics~\cite{handbook-la2017}, we extract relevant information from data of 46 students, model their progress, and present the selected results as graphical models to instructors (see \Cref{sec:methods}).

Our research goals are to examine how the graph models can support assessment and how we could improve them. We performed a study with 22 expert instructors who evaluated the graph models of selected students; \Cref{sec:results} presents the results. In \Cref{sec:discussion}, we compare the two modeling approaches and discuss their benefits, limitations, and practical implications for teaching and research. \Cref{sec:conclusions} concludes the paper and summarizes our contributions.

\section{Related Work in Assessment Models}
\label{sec:related-work}

Although in-depth assessment improves learning~\cite{weiss2016}, only a few studies have explored assessment models for security exercises. \Cref{subsec:related-work-cyber} reviews such work and explains how we differ. \Cref{subsec:related-work-other} discusses other models used in computing education research.

\subsection{Hands-on Cybersecurity Education}
\label{subsec:related-work-cyber}

Visualizations of student data and learning content are valuable in education~\cite{Firat2018}. Ošlejšek et al.~\cite{oslejsek2020} demonstrated this in the context of cybersecurity training. They proposed multiple visualizations to support the instructors' classroom awareness and student assessment. The visualizations display data of student interaction with the training environment, such as the submission of incorrect answers and help requests. The authors claim that visual models \enquote{should provide an overview as well as detailed per-trainee data.}

Andreolini et al.~\cite{andreolini2019framework} used directed graphs to model student progress in a security exercise. The vertices of the graphs represent the intermediate states of the exercise. The edges represent the actions that trigger a state transition. The graphs are generated automatically from a reference graph to assess trainee performance. A slight shortcoming is that the states' order is fixed, and paths not leading to the solution are disregarded. Braghin et al.~\cite{Braghin2020} proposed a follow-up: automated scoring metrics based on the reference graphs. However, the proposal is yet to be applied in practice.

Weiss et al.~\cite{weiss2016, weiss2017magazine} collected students' command histories from exercises in the EDURange platform. Using the data, they manually constructed graph models of student progress. The models revealed student approaches and misconceptions that would have been lost if the students were assessed only by the solution (in)correctness.

Mirkovic et al.~\cite{mirkovic2020, lepe2019} developed a system that assesses student progress in hands-on assignments in DETERlab and EDURange platforms. The system collects the input and output of the student's command line and matches the logs with pre-defined milestones (subgoals for the assignment). The system helps the instructors to monitor student learning and identify challenging concepts.

We extend the previous work by evaluating the models with instructors from multiple institutions. We also automate some manual aspects and extend the modeling capabilities to include solutions to partially ordered tasks. These improvements allow us to model a wider variety of exercises in multiple platforms.

\subsection{Other Areas of Computing Education}
\label{subsec:related-work-other}

Modeling formalisms applied in computing education include:
\begin{itemize}[leftmargin=5mm]
    \item \textit{Petri nets}~\cite{peterson1977petri}, which were used to model how students progressed through a study curriculum~\cite{handbook-edm9},
    \item \textit{Bayesian networks}~\cite{charniak1991bayesian, millan2010bayesian} to predict student attitudes and goals in a tutoring system~\cite{handbook-edm23} or test performance~\cite{handbook-edm29}, and
    \item \textit{Markov decision processes} to generate automated hints~\cite{handbook-edm33}.
\end{itemize}
While these studies used student data as input for statistical and machine learning methods, we construct visual models for teachers.

Piech et al.~\cite{piech2012modeling} captured and clustered temporal traces of student interactions with a compiler to study how students learn to program. They applied a hidden Markov model to the traces and visualized it as a state machine for the cluster. The models then predicted student performance. In our case, the exercise milestones are clearer and easier to define, though this approach could be applied as well.

Hooshyar et al.~\cite{Hooshyar2018} reviewed methods for modeling the players of educational games. They identified \enquote{\textit{data-driven approaches to conceptualizing log data}} as a promising research direction. They see a major challenge in determining actions that \enquote{\textit{represent key features of player performance.}} Our research attempts to address this problem in the domain of hands-on cybersecurity education.

\section{Study and Assessment Methods}
\label{sec:methods}

In this paper, we use the term \textit{exercise} to denote a set of assignments in which the students practice their cybersecurity skills. We host cybersecurity exercises in two interactive learning environments: \kypo~\cite{my-2021-FIE-KYPO-CSC} and \edurange~\cite{edurange_SIGCSE2015}. For each, we now describe the exercise content, participating students, and the process of generating the graph models from students' command-line data. Then, we detail the research methods for the graph models evaluation.

\subsection{Exercise Environment and Content}

In the \kypo\ environment, the students interact with virtual machines (VMs) in an emulated network to solve the exercise tasks. For this research, we used an exercise \locust~\cite{locust3302-gitlab} created within the Seminar on the Simulation of Cyber Attacks~\cite{svabensky2018kypolab}. Students assume the role of a cyber investigator who tracks a fictional hacker group. The students have to scan a suspicious server using \texttt{nmap}~\cite{nmap}, identify a vulnerable service, and exploit it using Metasploit~\cite{metasploit} to gain access. Then, they have to copy a private SSH key, crack its passphrase using John the Ripper (\texttt{john})~\cite{john}, and use it to access another host that stores secret documents.

For exercises in the \edurange\ platform, students use an SSH client to connect to one or more Linux VMs. To achieve variety, we chose an exercise called \filewrangler, which is entirely different from \locust. Students worked only with one VM to perform tasks such as finding hidden files, identifying file formats, and changing access permissions.

In both exercises, the tasks are also gamified in that students find text strings called \enquote{flags} by discovering secret files.

\subsection{Teaching Context and Student Data}

\kypo\ hosted the \locust\ exercise for 20 participants, undergraduates and advanced high school students, in a summer school held remotely in July 2020. During the two-hour training session, we recorded 2,382 commands submitted by the students. The data include full commands with their arguments and metadata, such as arbitrary student ID and timestamp~\cite{Svabensky2021dataset}. Since the students had limited time for the exercise, not all of them finished all the tasks.

\edurange\ deployed \filewrangler\ in a class of 26 students in an intermediate class in networking and network security in February 2020. The students were concentrating in computer science, and they had all taken an introductory course. Most students were familiar with the Linux command line. In total, 3,178 commands were recorded and analyzed from participants in this exercise.

For both exercises, the data were anonymized to protect the students' privacy. We received a waiver/approval from our respective institutions to process the data for this study.

\subsection{Model Generation from Student Data}

The collected student command logs are used to model progress through the exercise. We proposed two methods for generating graph models to support student assessment, which we describe below. Although the methods work in real-time as well, the scope of this study is post-exercise assessment. Therefore, all models were generated after all students finished their exercises.

\subsubsection{Trainee Graph}

In the first approach, the exercise author manually and iteratively creates a \textit{reference graph} that serves as a sample solution. Similarly to~\cite{andreolini2019framework}, the vertices of the reference graph represent the exercise subgoals, such as using the right tool. These states are desirable to reach. The directed edges represent the commands the student must execute to progress from one state to the next. The graphs are written in human-readable \textit{DOT} language~\cite{dot_language}.

\begin{figure}[!ht]
\centering
\includegraphics[width=0.95\linewidth]{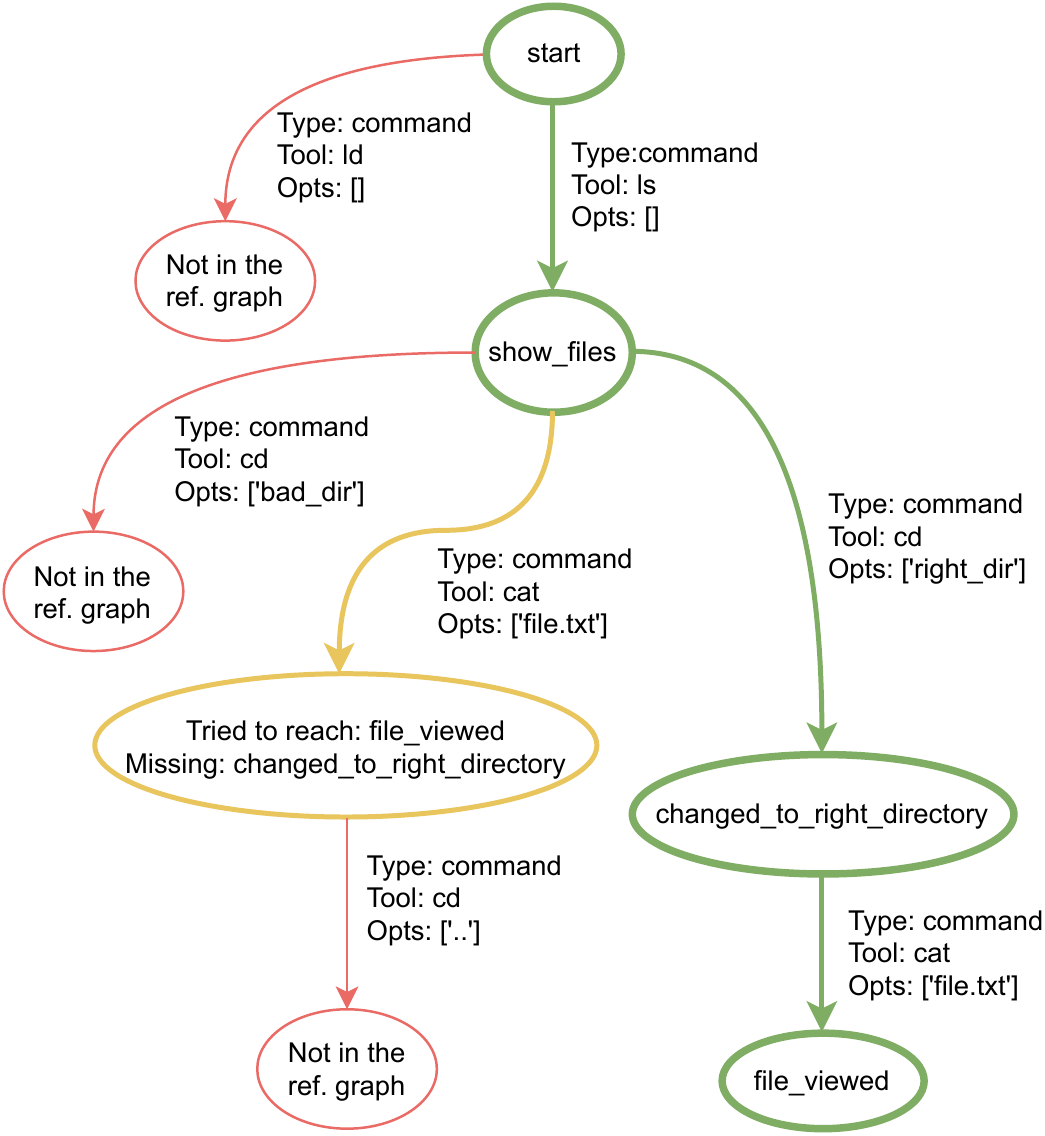}
\caption{A simplified \traineegraph. The green states and edges represent successful steps mapped to the reference graph. The red states and edges show actions that were likely erroneous or unnecessary. The yellow state and edge show an action with possibly missing prerequisites.}
\label{fig:trainee-graph}
\vspace*{-1mm}
\end{figure}

Then, each student's commands are automatically mapped to the reference graph using the NetworkX Python module~\cite{hagberg2008networkx} and visualized with Graphviz~\cite{Ellson03graphviz}. This results in our first model, a \traineegraph\ (see \Cref{fig:trainee-graph}). The pattern matching can map the student's command to any in-edge of any state. States can be reachable independently in an arbitrary order to allow modeling parallel tasks and skipping steps, or include prerequisite states to model a sequence of actions. On average, the graphs from our data included 39 states and 66 edges. For details about the graph generation, see~\cite[Sec. 4]{Chudovsky2020thesis}.

\subsubsection{Milestone Graph}

\begin{figure}[!ht]
\centering
\includegraphics[width=\linewidth]{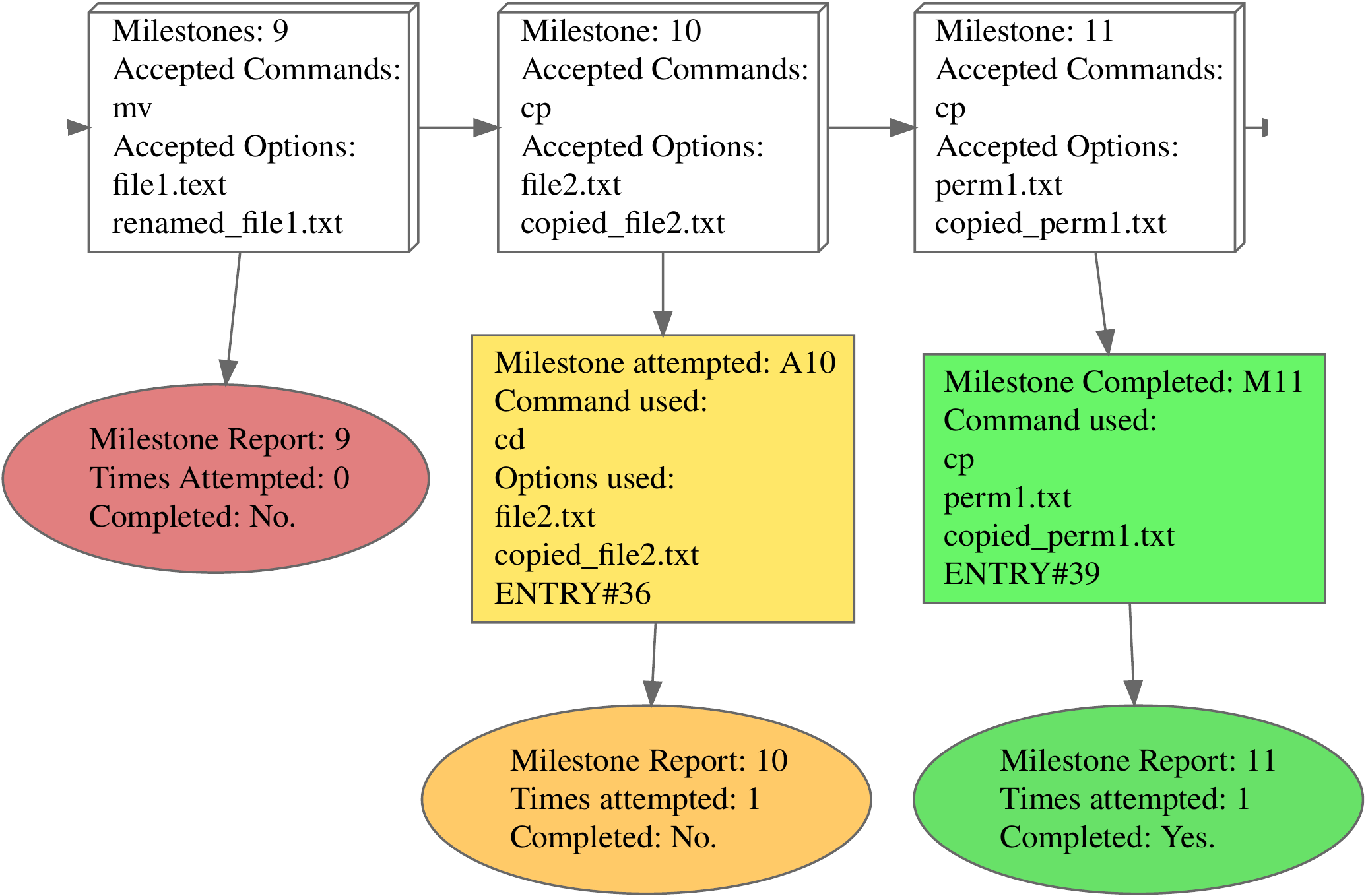}
\caption{\milestonegraph\ is composed of a chain of template nodes for each task. In this example, the student did not attempt the first of the three tasks, incorrectly attempted the second task (used \texttt{cd} instead of \texttt{cp}) but did not complete it, and completed the third task on their first try.}
\label{fig:milestone-graph}
\vspace*{-2mm}
\end{figure}

The other model, \milestonegraph~(see \Cref{fig:milestone-graph}), was constructed using a similar process but a different tool. The exercises are broken into tasks by the authors. For each task, specific regular expressions represent a milestone. Python scripts then read student Bash history input and output data~\cite{FSFBash2019}. Each line of the Bash history is split and checked against the regular expressions to find milestone attempts. If the line does not match all of the expressions, it is considered to be an unsuccessful attempt. The milestones are ordered by the author, but students do not need to complete them in that order.

The graph contains \textit{template nodes} that describe each milestone (the same for each student) and \textit{attempt nodes} connected to them that match the regular expressions for that milestone. Commands that do not match any milestone are not shown in the graph.

The generated files are processed by Graphviz. Successful attempts are drawn as green nodes, unsuccessful ones are yellow, and unattempted milestones are red. A summary node is appended to the chain based on whether the milestone was ultimately achieved. We also record the number of attempts per milestone (how many relevant commands the student tried). Sometimes matching unsuccessful commands with milestones is ambiguous. The milestones are ordered based on an expected path, which resolves ambiguities by associating an attempt with the earliest similar milestone. The commands' chronological order is encoded in the ENTRY numbers.

Both graph models are generic, and the tools for their creation accept input data from both learning environments. The tools would work with data from other environments too, as long as the structure of the input is preserved. This allowed us to compare the two models and would allow others to adopt or adapt them.

\subsection{Model Evaluation by Expert Instructors}

\begin{figure}[t]
\centering
\includegraphics[width=0.9\linewidth]{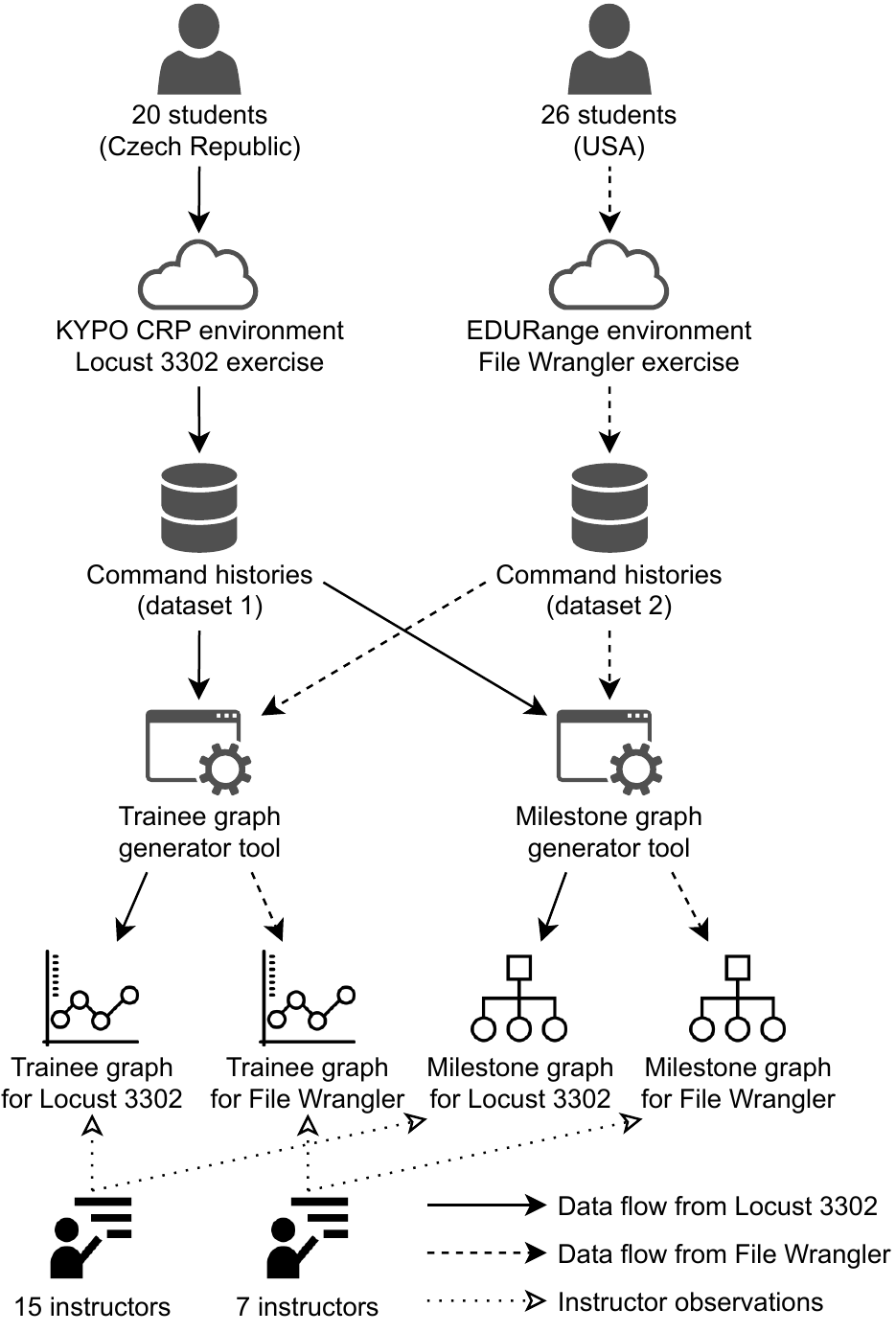}
\caption{Overview of the factorial design of the study.}
\label{fig:overview}
\end{figure}

\Cref{fig:overview} summarizes the setup of our study. After generating the graph models from student data, we selected four graphs of two representative students (one from \locust\ and the other from \filewrangler) who did well on the exercise but had some problems. We did not choose students who did very well or poorly because that would have been obvious in both models and would have yielded less information.

Then, we asked 40 experienced computing instructors to qualitatively evaluate the models. The instructors were our current or former co-workers from about a dozen different institutions, such as a public college, liberal arts college, and a research university. Most of them had experience teaching cybersecurity but were not familiar with the two exercises. In total, 22 instructors responded (15 for \locust\ and 7 for \filewrangler). They all participated voluntarily and were not incentivized.

Each instructor received an e-mail explaining our research goals and the following attachments:
\begin{itemize}[leftmargin=5mm]
    \item Briefing that familiarized them with the exercise content (either \locust\ or \filewrangler),
    \item PDF files with a \traineegraph\ and a \milestonegraph\ generated from the data of one student for each exercise,
    \item Short instructions on how to interpret the two graphs.
\end{itemize}
The instructors were asked to examine the graph models and then answer the following questions in an anonymous online survey:
\begin{itemize}[leftmargin=6mm]
    \item[Q1] How do you assess the student's progress based on the two graph models you received?
    \item[Q2] In which tasks was the student struggling? Please describe them specifically.
    \item[Q3] What feedback would you provide to the student so that his/her learning can improve?
    \item[Q4] How do you compare the two types of graph models?
    \item[Q5] On a scale from 1 (not at all) to 5 (very), how useful would the graphs be for your classes?
\end{itemize}

Before the actual study, we performed a pilot test among the paper authors and two other instructors and clarified the instructions and survey wording. After collecting the responses, three authors performed open coding~\cite{handbook-CER7} of the replies to qualitatively analyze what information did the graph models convey.

\section{Results of the Survey Evaluation}
\label{sec:results}

We now present the results for each survey question, along with quotes from various participants to illustrate the points they made.

\subsection*{Q1: Assessing Student Progress}

Question 1 asked the instructors to assess the student according to the graph models. Out of 22 instructors, 14 said the student progressed pretty well based on at least one of the graphs. They commented that the student made progress and demonstrated a growing understanding over time. One instructor praised the student for experimenting with different approaches.
\begin{quote}
Overall the student is making good progress, but there [are] a few Linux tasks that need to be reviewed and practiced.
\end{quote}

Only four instructors assessed the student as having struggled with the whole exercise. Next, six other instructors interpreted the graphs as showing disparate amounts of student progress. Specifically, they assessed the student better based on the \milestonegraph\ and worse based on the \traineegraph. The reason is that the \traineegraph\ shows every student command-line entry, including a lot of trial and error, while the \milestonegraph\ omits student entries that do not match any milestone.
\begin{quote}
The trainee graph makes the student appear to have fewer skills than the milestone graph -- their struggle is much more visually pronounced in that representation.
\end{quote}

The graphs were different enough that, surprisingly, two instructors thought they came from different students (even though the study instructions said both graphs show the same student).

\subsection*{Q2: Identifying Difficulties}

Question 2 examined whether the instructors could identify from the graphs which parts of the exercise were problematic for the student, so that they could intervene. Most of the instructors were able to do that. For \locust, 10 out of 15 correctly identified at least one of the areas where the student struggled. For \filewrangler, 6 out of 7 identified the problem areas. Therefore, both graphs fulfilled the intended use case, although there is room for improvement.

\subsection*{Q3: Providing Feedback to the Student}

Question 3 asked the instructors how they would intervene after they identified where the student struggled. Our goal was to understand how the graphs could be used for in-class feedback.

From the pedagogical point of view, the instructors' feedback to the student differed a lot. Some suggested a direct approach, such as explaining the problem, the correct solution to it, and why the student's attempts were incorrect. They would also provide a tutorial or an example of the tool the student was struggling with.
\begin{quote}
The student needs feedback on how to better use the [Linux] shell and Metasploit console.
\end{quote}

Some instructors opted for a more indirect approach, such as suggesting to the student to find and learn what the commands do, review and understand their syntax, and read manual pages.
\begin{quote}
It is not uncommon to do trial and error on the Linux command line. But after the first failed attempt -- go to \texttt{man} [pages].
\end{quote}
They also emphasized the need to thoroughly read the task assignment and understanding it before starting to type commands.

Other instructors focused on affective feedback, such as encouraging the student to keep trying, praising them for their effort, and inviting them to ask the instructor for help.

All types of feedback were reasonable, given the information the instructors had. Three instructors noted that without the full assignment, it was hard to distinguish conceptual misunderstandings from tool-specific issues, which is a slight limitation of the survey.
\begin{quote}
As a participant in this study who isn't familiar with the example assignment, it's hard for me to distinguish high-level misconceptions (\enquote{I don't understand what \texttt{john} does, abstractly}) with low-level ones (\enquote{I know what \texttt{john} does but don't understand its command line arguments/syntax}). I'd probably focus in on specific learning goals and ask them about \texttt{john} and Metasploit and see what they do understand.
\end{quote}

\subsection*{Q4: Comparing the Two Graph Models}

Question 4 asked the instructors to compare the two graphs, and the vast majority agreed on the strengths and weaknesses of both. The \traineegraph\ was more detailed, which is a double-edged sword. On the one hand, it gives deeper insight into the student's work, including their used commands, problems, and solution attempts. On the other hand, the graph is difficult to interpret, and working with it is more time-consuming.
\begin{quote}
Trainee graph has more details, but, as a consequence, it is hard to read. It was much easier for me to understand and work with the Milestone graph. Nevertheless, Trainee graph shows [\dots] the wrong paths and gives the context unavailable in the Milestone graph (e.g., completely wrong directions).
\end{quote}

Still, instructors found the \traineegraph\ useful for evaluating and improving their exercise design.
\begin{quote}
Trainee graph lets me envision the temporal process of the student struggling, see where they got stuck, see where the design of my assignment maybe led them astray. If most of my students have similar graphs, that tells me a lot about which parts of my assignment were tricky, especially if most of them moved on past that challenge point (or didn't), whether I was reasonable in asking them to figure something out.
\end{quote}

The key reported weaknesses of the \traineegraph\ were that it shows any deviation from the reference solution as a potential error, making it difficult to detect unexpected solutions of students. It also becomes complicated with the growing number of commands and is not 100\% colorblind-friendly.

The instructors strongly agreed that \milestonegraph\ is easier to read; only two instructors found it difficult to interpret. Although it omits some details, it is simpler to work with. As a result, it provides a better quick overview (a \enquote{\textit{summarized breakdown}}) of student actions and is more suitable for batch assessment.
\begin{quote}
Milestone graph is much easier to assess an individual student's progress quickly, especially when grading many students.
\end{quote}

Since it also captures the attempted and completed task milestones (instead of states), it quickly shows what the student can or cannot do. This translates more directly to skill assessment.
\begin{quote}
Milestone graph feels more useful as a record of the student's skills and development. [I would use it] for providing learning-goal based formative feedback to my students.
\end{quote}

\subsection*{Q5: Usefulness Rating}

Finally, the instructors rated how useful they thought the graphs would be in their classes on a scale of 1 (not at all) to 5 (very). Some responded that their hands-on classes might not fit the structure needed for generating this type of data. However, the average score across 21 responses (one instructor did not answer) was 3.57 out of 5. The median and the mode was 4, which means they considered it beneficial but not perfect. One respondent noted that the payoff of automated solutions like these increases as class size grows.

\section{Discussion}
\label{sec:discussion}

This section summarizes the lessons learned from the survey, discusses its limitations, and, based on that, proposes future work.

\subsection{Summary of the Results}

Developing two graph models proved useful, since their evaluation elicited different perspectives. Most instructors interpreted each graph effectively, and they also identified strengths, weaknesses, and use cases for each graph. They reported that the \traineegraph\ provided more detail by mapping command history as it happened. This can help review and improve exercise design, as well as discover unexpected solutions by examining the red and yellow elements. In contrast, the \milestonegraph\ showed key stages of student's progress and was easier to read. It is better in providing a quick overview, especially in time-critical situations, and supports assessment based on learning goals. Overall, the instructors found both models useful and novel but also noted their shortcomings, mainly the time required for large classes.
\begin{quote}
I see the great potential in visualizing commands for better analysis of students' thinking. [The models] might provide better insight into their work. However, in their current form, they are far from ideal. I would need to analyze [the models] one by one and formulate suggestions for the students independently.
\end{quote}

Another participant commented that the graphs could be shown to students as feedback. This could highlight common misconceptions or missed learning objectives across an entire class.
\begin{quote}
I like the idea of graphically summarizing the student experience [\ldots]. I can imagine displaying a bunch of graphs for all students in a class side-by-side and having the common problems jump out visually. This would help the instructor know what to emphasize in the next class session.
\end{quote}

\subsection{Implications for Teaching Practice}

Since both graphs visualize the task subgoals and student attempts, the graphs could be used for the following educational use cases:
\begin{itemize}[leftmargin=5mm]
    \item identify high- or low-performing students in class;
    \item identify successes and struggles of a specific student;
    \item assess students, both formatively and summatively; and
    \item give each student their own graph to reflect on their approaches, self-evaluate their learning process, and identify problems in the steps they chose to solve the exercise.
\end{itemize}

A key feature is that these use cases apply to both in-person and distance education. Supporting remote assessment is crucial when the instructor has limited access to what the students are doing. Moreover, since the graphs can be generated in real-time, they can be used for in-class interventions, not only post-exercise feedback as in this study. This feature becomes especially relevant if the graph generation is incorporated into the learning environment. Last but not least, the graph models are applicable not only in the cybersecurity domain, but generalize to any learning exercise that can be represented by a series of actions.

\subsection{Addressing the Limitations}

The evaluation also revealed the limitations of the graph models. In complex exercises, the reference graph or milestone definitions that enable model generation can be incomplete. Thus, a correct but unexpected student solution could be marked as erroneous. However, an instructor can gradually update the definitions and generate new graphs. 

Another limitation is that the graphs do not scale for sequences of hundreds of commands. This can be resolved by splitting complex exercises into sections, or implementing the graphs as interactive visualizations with collapsible parts and filters.

Some may consider a limitation that the tools are primarily designed for command-line exercises. However, command line interface is important in  practice, and the tools would also work with a variety of log files, e.g., webserver or database query logs. Relevant exercises include not only the cybersecurity domain, but also programming, operating systems, and networking. Given a reference graph or regular expressions tailored to these data sources, the graphs can be extended to display many types of student activity.

Regarding the study validity, we received survey responses from 22 instructors out of the 40 asked. Although this number is not high enough to allow generalizing the results, the sample represents instructors with different backgrounds. While selection bias may be present, since we asked mostly our current or former colleagues, the multi-institutional study framework should mitigate the bias. The final limitation is that although we had data from 46 students, we selected only two of them. The reason is that since we did not incentivize the instructors who participated in the survey, we did not want to take too much of their time by asking them to evaluate more graphs. Nevertheless, we selected representative students to illustrate various aspects that appeared in other graphs as well.

\subsection{Future Work}

Future studies can evaluate the effectiveness of the graph modeling approach with students in real-time. It can be interesting to examine whether students would find the information in the graphs useful. Another follow-up study can be more longitudinal, investigating whether the performance of a single student as displayed by the graphs improves over several training sessions.

Future research can also incorporate machine learning methods. Clustering can group students based on their performance. This would scale to large classes and save the instructors' time because they would not need to examine each student, only the representative of a cluster, and provide feedback applicable to the whole cluster. Alternatively, classification can be used to assess students automatically and even live during the exercise, indicating their skill level or the correctness of their actions. This solution would address the suggestion of one of the study participants:
\begin{quote}
I could see having both these graphs being potentially quite useful, especially if it updated live as my students worked, allowing me to catch common areas of concern.
\end{quote}

\section{Conclusions}
\label{sec:conclusions}

Assessment of learning is crucial to provide instructors with classroom situational awareness, identify students' strengths and shortcomings, and help students learn. This work proposed two methods for assessing student progress in hands-on exercises. The methods visualize and contextualize command history logs, which are very hard to process manually in their raw form. One method provides a quick summary; the other complements it with a detailed view. Together, they improve understanding of students' approaches to learning and represent a faster form of feedback than traditional post-homework assessment. Another strength of this collaborative research is that by giving the instructors two models to compare, they precisely formulated what worked and what did not in each.

We implemented the methods as open-source tools and used them to generate 46 graph models from authentic in-class data: 5,560 commands submitted by students at two universities (approx. 121 lines per student of minimally formatted text logs). The source code and data are available in a public repository~\cite{dataset} to support their adoption by other instructors and researchers.

The resulting graph models were evaluated by 22 instructors from various institutions. Qualitative analysis of their responses revealed strengths, weaknesses, and applications of the two proposed methods in assessment. They can highlight student skills, provide a basis for classroom interventions, and reveal issues in exercise design. Since the methods are generic, they are applicable in other learning environments and exercises. Moreover, they can be applied outside of the cybersecurity domain to enhance assessment in other computing classes that capture student interaction.

\begin{acks}
The researchers from Masaryk University were supported by \grantsponsor{ERDF}{ERDF}{} project \textit{CyberSecurity, CyberCrime and Critical Information Infrastructures Center of Excellence} (No. \grantnum{ERDF}{CZ.02.1.01/0.0/0.0/16\_019/0000822}). Part of this paper is based upon work supported by the National Science Foundation under grant numbers 1723705 and 1723714. Finally, we thank the instructors who participated in this study, as well as the SIGCSE conference reviewers who provided constructive feedback that helped improve the final version of paper.
\end{acks}

\balance
\bibliographystyle{ACM-Reference-Format}
\bibliography{references}


\begin{thebibliography}{37}


\ifx \showCODEN    \undefined \def \showCODEN     #1{\unskip}     \fi
\ifx \showDOI      \undefined \def \showDOI       #1{#1}\fi
\ifx \showISBNx    \undefined \def \showISBNx     #1{\unskip}     \fi
\ifx \showISBNxiii \undefined \def \showISBNxiii  #1{\unskip}     \fi
\ifx \showISSN     \undefined \def \showISSN      #1{\unskip}     \fi
\ifx \showLCCN     \undefined \def \showLCCN      #1{\unskip}     \fi
\ifx \shownote     \undefined \def \shownote      #1{#1}          \fi
\ifx \showarticletitle \undefined \def \showarticletitle #1{#1}   \fi
\ifx \showURL      \undefined \def \showURL       {\relax}        \fi
\providecommand\bibfield[2]{#2}
\providecommand\bibinfo[2]{#2}
\providecommand\natexlab[1]{#1}
\providecommand\showeprint[2][]{arXiv:#2}

\bibitem[\protect\citeauthoryear{Andreolini, Colacino, Colajanni, and
  Marchetti}{Andreolini et~al\mbox{.}}{2019}]%
        {andreolini2019framework}
\bibfield{author}{\bibinfo{person}{Mauro Andreolini}, \bibinfo{person}{Vincenzo
  Colacino}, \bibinfo{person}{Michele Colajanni}, {and} \bibinfo{person}{Mirco
  Marchetti}.} \bibinfo{year}{2019}\natexlab{}.
\newblock \showarticletitle{{A Framework for the Evaluation of Trainee
  Performance in Cyber Range Exercises}}.
\newblock \bibinfo{journal}{\emph{Mobile Networks and Applications}}
  \bibinfo{volume}{25} (\bibinfo{year}{2019}).
\newblock
\urldef\tempurl%
\url{https://doi.org/10.1007/s11036-019-01442-0}
\showDOI{\tempurl}


\bibitem[\protect\citeauthoryear{Arroyo, Cooper, Burleson, and Woolf}{Arroyo
  et~al\mbox{.}}{2010}]%
        {handbook-edm23}
\bibfield{author}{\bibinfo{person}{Ivon Arroyo}, \bibinfo{person}{David~G.
  Cooper}, \bibinfo{person}{Winslow Burleson}, {and}
  \bibinfo{person}{Beverly~P. Woolf}.} \bibinfo{year}{2010}\natexlab{}.
\newblock \showarticletitle{Bayesian Networks and Linear Regression Models of
  Students’ Goals, Moods, and Emotions}.
\newblock In \bibinfo{booktitle}{\emph{Handbook of educational data mining}},
  \bibfield{editor}{\bibinfo{person}{Cristobal Romero},
  \bibinfo{person}{Sebastian Ventura}, \bibinfo{person}{Mykola Pechenizkiy},
  {and} \bibinfo{person}{Ryan~S.J.d. Baker}} (Eds.). \bibinfo{publisher}{CRC
  Press}, \bibinfo{address}{Boca Raton, FL, USA}, Chapter~23,
  \bibinfo{pages}{323--338}.
\newblock
\showISBNx{978-1-4398-0458-2}
\urldef\tempurl%
\url{https://doi.org/10.1201/b10274}
\showDOI{\tempurl}


\bibitem[\protect\citeauthoryear{Barnes, Stamper, and Croy}{Barnes
  et~al\mbox{.}}{2010}]%
        {handbook-edm33}
\bibfield{author}{\bibinfo{person}{Tiffany Barnes}, \bibinfo{person}{John
  Stamper}, {and} \bibinfo{person}{Marvin Croy}.}
  \bibinfo{year}{2010}\natexlab{}.
\newblock \showarticletitle{Using Markov Decision Processes for Automatic Hint
  Generation}.
\newblock In \bibinfo{booktitle}{\emph{Handbook of educational data mining}},
  \bibfield{editor}{\bibinfo{person}{Cristobal Romero},
  \bibinfo{person}{Sebastian Ventura}, \bibinfo{person}{Mykola Pechenizkiy},
  {and} \bibinfo{person}{Ryan~S.J.d. Baker}} (Eds.). \bibinfo{publisher}{CRC
  Press}, \bibinfo{address}{Boca Raton, FL, USA}, Chapter~33,
  \bibinfo{pages}{467--480}.
\newblock
\showISBNx{978-1-4398-0458-2}
\urldef\tempurl%
\url{https://doi.org/10.1201/b10274}
\showDOI{\tempurl}


\bibitem[\protect\citeauthoryear{Braghin, Cimato, Damiani, Frati, Riccobene,
  and Astaneh}{Braghin et~al\mbox{.}}{2020}]%
        {Braghin2020}
\bibfield{author}{\bibinfo{person}{Chiara Braghin}, \bibinfo{person}{Stelvio
  Cimato}, \bibinfo{person}{Ernesto Damiani}, \bibinfo{person}{Fulvio Frati},
  \bibinfo{person}{Elvinia Riccobene}, {and} \bibinfo{person}{Sadegh Astaneh}.}
  \bibinfo{year}{2020}\natexlab{}.
\newblock \showarticletitle{Towards the Monitoring and Evaluation of Trainees'
  Activities in Cyber Ranges}. In \bibinfo{booktitle}{\emph{Model-driven
  Simulation and Training Environments for Cybersecurity}},
  \bibfield{editor}{\bibinfo{person}{George Hatzivasilis} {and}
  \bibinfo{person}{Sotiris Ioannidis}} (Eds.). \bibinfo{publisher}{Springer
  International Publishing}, \bibinfo{address}{Cham}, \bibinfo{pages}{79--91}.
\newblock
\showISBNx{978-3-030-62433-0}
\urldef\tempurl%
\url{https://doi.org/10.1007/978-3-030-62433-0_5}
\showDOI{\tempurl}


\bibitem[\protect\citeauthoryear{Charniak}{Charniak}{1991}]%
        {charniak1991bayesian}
\bibfield{author}{\bibinfo{person}{Eugene Charniak}.}
  \bibinfo{year}{1991}\natexlab{}.
\newblock \showarticletitle{{Bayesian networks without tears}}.
\newblock \bibinfo{journal}{\emph{AI magazine}} \bibinfo{volume}{12},
  \bibinfo{number}{4} (\bibinfo{year}{1991}), \bibinfo{pages}{50--50}.
\newblock
\urldef\tempurl%
\url{https://doi.org/10.1609/aimag.v12i4.918}
\showDOI{\tempurl}


\bibitem[\protect\citeauthoryear{Chudovský}{Chudovský}{2020}]%
        {Chudovsky2020thesis}
\bibfield{author}{\bibinfo{person}{Radoslav Chudovský}.}
  \bibinfo{year}{2020}\natexlab{}.
\newblock \emph{\bibinfo{title}{{Modeling Progress Through Cybersecurity
  Training Using Command Histories}}}.
\newblock Bachelor's Thesis. \bibinfo{school}{Masaryk University, Faculty of
  Informatics}.
\newblock
\urldef\tempurl%
\url{https://is.muni.cz/th/hpykg/?lang=en}
\showURL{%
\tempurl}


\bibitem[\protect\citeauthoryear{Ellson et~al\mbox{.}}{Ellson
  et~al\mbox{.}}{2021}]%
        {dot_language}
\bibfield{author}{\bibinfo{person}{John Ellson} {et~al\mbox{.}}}
  \bibinfo{year}{2021}\natexlab{}.
\newblock \bibinfo{booktitle}{\emph{{DOT: Graph description language}}}.
\newblock {GraphViz}.
\newblock
\urldef\tempurl%
\url{https://www.graphviz.org/doc/info/lang.html/}
\showURL{%
Retrieved November 19, 2021 from \tempurl}


\bibitem[\protect\citeauthoryear{Ellson, Gansner, Koutsofios, North, and
  Woodhull}{Ellson et~al\mbox{.}}{2004}]%
        {Ellson03graphviz}
\bibfield{author}{\bibinfo{person}{John Ellson}, \bibinfo{person}{Emden~R.
  Gansner}, \bibinfo{person}{Eleftherios Koutsofios},
  \bibinfo{person}{Stephen~C. North}, {and} \bibinfo{person}{Gordon Woodhull}.}
  \bibinfo{year}{2004}\natexlab{}.
\newblock \bibinfo{booktitle}{\emph{{Graphviz and Dynagraph --- Static and
  Dynamic Graph Drawing Tools}}}.
\newblock \bibinfo{publisher}{Springer Berlin Heidelberg},
  \bibinfo{address}{Berlin, Heidelberg}, \bibinfo{pages}{127--148}.
\newblock
\showISBNx{978-3-642-18638-7}
\urldef\tempurl%
\url{https://doi.org/10.1007/978-3-642-18638-7_6}
\showDOI{\tempurl}


\bibitem[\protect\citeauthoryear{Firat and Laramee}{Firat and Laramee}{2018}]%
        {Firat2018}
\bibfield{author}{\bibinfo{person}{Elif~E. Firat} {and}
  \bibinfo{person}{Robert~S. Laramee}.} \bibinfo{year}{2018}\natexlab{}.
\newblock \showarticletitle{Towards a Survey of Interactive Visualization for
  Education}. In \bibinfo{booktitle}{\emph{Proceedings of the Conference on
  Computer Graphics \& Visual Computing}} \emph{(\bibinfo{series}{CGVC '18})}.
  \bibinfo{publisher}{Eurographics Association}, \bibinfo{address}{Goslar,
  DEU}, \bibinfo{pages}{91–101}.
\newblock
\urldef\tempurl%
\url{https://doi.org/10.2312/cgvc.20181211}
\showDOI{\tempurl}


\bibitem[\protect\citeauthoryear{Force}{Force}{2020}]%
        {cc2020}
\bibfield{author}{\bibinfo{person}{CC2020~Task Force}.}
  \bibinfo{year}{2020}\natexlab{}.
\newblock \bibinfo{booktitle}{\emph{Computing Curricula 2020: Paradigms for
  Global Computing Education}}.
\newblock \bibinfo{publisher}{Association for Computing Machinery},
  \bibinfo{address}{New York, NY, USA}.
\newblock
\showISBNx{978-1-4503-9059-0}
\urldef\tempurl%
\url{https://doi.org/10.1145/3467967}
\showDOI{\tempurl}


\bibitem[\protect\citeauthoryear{Guzdial and du~Boulay}{Guzdial and
  du~Boulay}{2019}]%
        {handbook-CER1}
\bibfield{author}{\bibinfo{person}{Mark Guzdial} {and}
  \bibinfo{person}{Benedict du Boulay}.} \bibinfo{year}{2019}\natexlab{}.
\newblock \showarticletitle{The History of Computing Education Research}.
\newblock In \bibinfo{booktitle}{\emph{The Cambridge Handbook of Computing
  Education Research}}, \bibfield{editor}{\bibinfo{person}{Sally~A Fincher}
  {and} \bibinfo{person}{Anthony~V Robins}} (Eds.).
  \bibinfo{publisher}{Cambridge University Press}, \bibinfo{address}{Cambridge,
  United Kingdom}, Chapter~1, \bibinfo{pages}{11--39}.
\newblock
\showISBNx{978-1-108-72189-9}
\urldef\tempurl%
\url{https://doi.org/10.1017/9781108654555}
\showDOI{\tempurl}


\bibitem[\protect\citeauthoryear{Hagberg, Schult, and Swart}{Hagberg
  et~al\mbox{.}}{2008}]%
        {hagberg2008networkx}
\bibfield{author}{\bibinfo{person}{Aric~A. Hagberg}, \bibinfo{person}{Daniel~A.
  Schult}, {and} \bibinfo{person}{Pieter~J. Swart}.}
  \bibinfo{year}{2008}\natexlab{}.
\newblock \showarticletitle{{Exploring Network Structure, Dynamics, and
  Function using NetworkX}}. In \bibinfo{booktitle}{\emph{Proceedings of the
  7th Python in Science Conference}}, \bibfield{editor}{\bibinfo{person}{Ga\"el
  Varoquaux}, \bibinfo{person}{Travis Vaught}, {and} \bibinfo{person}{Jarrod
  Millman}} (Eds.). \bibinfo{publisher}{SciPy Conferences},
  \bibinfo{address}{Pasadena, CA, USA}, \bibinfo{pages}{11--15}.
\newblock
\urldef\tempurl%
\url{https://www.osti.gov/biblio/960616}
\showURL{%
\tempurl}


\bibitem[\protect\citeauthoryear{Hooshyar, Yousefi, and Lim}{Hooshyar
  et~al\mbox{.}}{2018}]%
        {Hooshyar2018}
\bibfield{author}{\bibinfo{person}{Danial Hooshyar}, \bibinfo{person}{Moslem
  Yousefi}, {and} \bibinfo{person}{Heuiseok Lim}.}
  \bibinfo{year}{2018}\natexlab{}.
\newblock \showarticletitle{Data-Driven Approaches to Game Player Modeling: A
  Systematic Literature Review}.
\newblock \bibinfo{journal}{\emph{ACM Comput. Surv.}} \bibinfo{volume}{50},
  \bibinfo{number}{6}, Article \bibinfo{articleno}{90} (\bibinfo{date}{Jan.}
  \bibinfo{year}{2018}), \bibinfo{numpages}{19}~pages.
\newblock
\showISSN{0360-0300}
\urldef\tempurl%
\url{https://doi.org/10.1145/3145814}
\showDOI{\tempurl}


\bibitem[\protect\citeauthoryear{Laboratory}{Laboratory}{2021}]%
        {locust3302-gitlab}
\bibfield{author}{\bibinfo{person}{Cybersecurity Laboratory}.}
  \bibinfo{year}{2021}\natexlab{}.
\newblock \bibinfo{booktitle}{\emph{{Locust 3302}}}.
\newblock Masaryk University.
\newblock
\urldef\tempurl%
\url{https://gitlab.ics.muni.cz/muni-kypo-trainings/games/locust-3302}
\showURL{%
Retrieved November 19, 2021 from \tempurl}


\bibitem[\protect\citeauthoryear{Lancaster, Robins, and Fincher}{Lancaster
  et~al\mbox{.}}{2019}]%
        {handbook-CER14}
\bibfield{author}{\bibinfo{person}{Thomas Lancaster},
  \bibinfo{person}{Anthony~V. Robins}, {and} \bibinfo{person}{Sally~A.
  Fincher}.} \bibinfo{year}{2019}\natexlab{}.
\newblock \showarticletitle{Assessment and Plagiarism}.
\newblock In \bibinfo{booktitle}{\emph{The Cambridge Handbook of Computing
  Education Research}}, \bibfield{editor}{\bibinfo{person}{Sally~A Fincher}
  {and} \bibinfo{person}{Anthony~V Robins}} (Eds.).
  \bibinfo{publisher}{Cambridge University Press}, \bibinfo{address}{Cambridge,
  United Kingdom}, Chapter~14, \bibinfo{pages}{414--444}.
\newblock
\showISBNx{978-1-108-72189-9}
\urldef\tempurl%
\url{https://doi.org/10.1017/9781108654555}
\showDOI{\tempurl}


\bibitem[\protect\citeauthoryear{Lang, Siemens, Wise, and Gašević}{Lang
  et~al\mbox{.}}{2017}]%
        {handbook-la2017}
\bibfield{editor}{\bibinfo{person}{Charles Lang}, \bibinfo{person}{George
  Siemens}, \bibinfo{person}{Alyssa Wise}, {and} \bibinfo{person}{Dragan
  Gašević}} (Eds.). \bibinfo{year}{2017}\natexlab{}.
\newblock \bibinfo{booktitle}{\emph{Handbook of Learning Analytics}
  (\bibinfo{edition}{1st} ed.)}.
\newblock \bibinfo{publisher}{Society for Learning Analytics Research (SoLAR)}.
\newblock
\showISBNx{978-0-9952408-0-3}
\urldef\tempurl%
\url{https://doi.org/10.18608/hla17}
\showDOI{\tempurl}


\bibitem[\protect\citeauthoryear{{Lepe}, {Aggarwal}, {Mirkovic}, {Mache},
  {Weiss}, and {Weinmann}}{{Lepe} et~al\mbox{.}}{2019}]%
        {lepe2019}
\bibfield{author}{\bibinfo{person}{Paul {Lepe}}, \bibinfo{person}{Aashray
  {Aggarwal}}, \bibinfo{person}{Jelena {Mirkovic}}, \bibinfo{person}{Jens
  {Mache}}, \bibinfo{person}{Richard {Weiss}}, {and} \bibinfo{person}{David
  {Weinmann}}.} \bibinfo{year}{2019}\natexlab{}.
\newblock \showarticletitle{Measuring Student Learning On Network Testbeds}. In
  \bibinfo{booktitle}{\emph{2019 IEEE 27th International Conference on Network
  Protocols (ICNP)}}. \bibinfo{publisher}{IEEE}, \bibinfo{address}{New York,
  NY, USA}, \bibinfo{pages}{1--2}.
\newblock
\showISSN{1092-1648}
\urldef\tempurl%
\url{https://doi.org/10.1109/ICNP.2019.8888101}
\showDOI{\tempurl}


\bibitem[\protect\citeauthoryear{Lyon}{Lyon}{2021}]%
        {nmap}
\bibfield{author}{\bibinfo{person}{Gordon Lyon}.}
  \bibinfo{year}{2021}\natexlab{}.
\newblock \bibinfo{booktitle}{\emph{{Nmap Network Scanning}}}.
\newblock Nmap.
\newblock
\urldef\tempurl%
\url{https://nmap.org/book/man.html}
\showURL{%
Retrieved November 19, 2021 from \tempurl}


\bibitem[\protect\citeauthoryear{Mill{\'a}n, Loboda, and
  P{\'e}rez-De-La-Cruz}{Mill{\'a}n et~al\mbox{.}}{2010}]%
        {millan2010bayesian}
\bibfield{author}{\bibinfo{person}{Eva Mill{\'a}n}, \bibinfo{person}{Tomasz
  Loboda}, {and} \bibinfo{person}{Jose~Luis P{\'e}rez-De-La-Cruz}.}
  \bibinfo{year}{2010}\natexlab{}.
\newblock \showarticletitle{{Bayesian networks for student model engineering}}.
\newblock \bibinfo{journal}{\emph{Computers \& Education}}
  \bibinfo{volume}{55}, \bibinfo{number}{4} (\bibinfo{year}{2010}),
  \bibinfo{pages}{1663--1683}.
\newblock
\showISSN{0360-1315}
\urldef\tempurl%
\url{https://doi.org/10.1016/j.compedu.2010.07.010}
\showDOI{\tempurl}


\bibitem[\protect\citeauthoryear{Mirkovic, Aggarwal, Weinman, Lepe, Mache, and
  Weiss}{Mirkovic et~al\mbox{.}}{2020}]%
        {mirkovic2020}
\bibfield{author}{\bibinfo{person}{Jelena Mirkovic}, \bibinfo{person}{Aashray
  Aggarwal}, \bibinfo{person}{David Weinman}, \bibinfo{person}{Paul Lepe},
  \bibinfo{person}{Jens Mache}, {and} \bibinfo{person}{Richard Weiss}.}
  \bibinfo{year}{2020}\natexlab{}.
\newblock \showarticletitle{{Using Terminal Histories to Monitor Student
  Progress on Hands-on Exercises}}. In \bibinfo{booktitle}{\emph{Proceedings of
  the 51st ACM Technical Symposium on Computer Science Education}}
  \emph{(\bibinfo{series}{SIGCSE ’20})}. \bibinfo{publisher}{ACM},
  \bibinfo{address}{New York, NY, USA}, \bibinfo{pages}{866–872}.
\newblock
\showISBNx{9781450367936}
\urldef\tempurl%
\url{https://doi.org/10.1145/3328778.3366935}
\showDOI{\tempurl}


\bibitem[\protect\citeauthoryear{Openwall}{Openwall}{2021}]%
        {john}
\bibfield{author}{\bibinfo{person}{Openwall}.} \bibinfo{year}{2021}\natexlab{}.
\newblock \bibinfo{booktitle}{\emph{{John the Ripper password cracker}}}.
\newblock
\urldef\tempurl%
\url{https://www.openwall.com/john/}
\showURL{%
Retrieved November 19, 2021 from \tempurl}


\bibitem[\protect\citeauthoryear{Ošlejšek, Rusňák, Burská, Švábenský,
  Vykopal, and Čegan}{Ošlejšek et~al\mbox{.}}{2021}]%
        {oslejsek2020}
\bibfield{author}{\bibinfo{person}{Radek Ošlejšek}, \bibinfo{person}{Vít
  Rusňák}, \bibinfo{person}{Karolína Burská}, \bibinfo{person}{Valdemar
  Švábenský}, \bibinfo{person}{Jan Vykopal}, {and} \bibinfo{person}{Jakub
  Čegan}.} \bibinfo{year}{2021}\natexlab{}.
\newblock \showarticletitle{Conceptual Model of Visual Analytics for Hands-on
  Cybersecurity Training}.
\newblock \bibinfo{journal}{\emph{IEEE Transactions on Visualization and
  Computer Graphics}} \bibinfo{volume}{27}, \bibinfo{number}{8}
  (\bibinfo{year}{2021}), \bibinfo{pages}{3425--3437}.
\newblock
\urldef\tempurl%
\url{https://doi.org/10.1109/TVCG.2020.2977336}
\showDOI{\tempurl}


\bibitem[\protect\citeauthoryear{Pardos, Heffernan, Anderson, and
  Heffernan}{Pardos et~al\mbox{.}}{2010}]%
        {handbook-edm29}
\bibfield{author}{\bibinfo{person}{Zachary~A. Pardos}, \bibinfo{person}{Neil~T.
  Heffernan}, \bibinfo{person}{Brigham~S. Anderson}, {and}
  \bibinfo{person}{Cristina~L. Heffernan}.} \bibinfo{year}{2010}\natexlab{}.
\newblock \showarticletitle{Using Fine-Grained Skill Models to Fit Student
  Performance with Bayesian Networks}.
\newblock In \bibinfo{booktitle}{\emph{Handbook of educational data mining}},
  \bibfield{editor}{\bibinfo{person}{Cristobal Romero},
  \bibinfo{person}{Sebastian Ventura}, \bibinfo{person}{Mykola Pechenizkiy},
  {and} \bibinfo{person}{Ryan~S.J.d. Baker}} (Eds.). \bibinfo{publisher}{CRC
  Press}, \bibinfo{address}{Boca Raton, FL, USA}, Chapter~29,
  \bibinfo{pages}{417--426}.
\newblock
\showISBNx{978-1-4398-0458-2}
\urldef\tempurl%
\url{https://doi.org/10.1201/b10274}
\showDOI{\tempurl}


\bibitem[\protect\citeauthoryear{Peterson}{Peterson}{1977}]%
        {peterson1977petri}
\bibfield{author}{\bibinfo{person}{James~L. Peterson}.}
  \bibinfo{year}{1977}\natexlab{}.
\newblock \showarticletitle{{Petri Nets}}.
\newblock \bibinfo{journal}{\emph{ACM Comput. Surv.}} \bibinfo{volume}{9},
  \bibinfo{number}{3} (\bibinfo{date}{Sept.} \bibinfo{year}{1977}),
  \bibinfo{pages}{223--252}.
\newblock
\showISSN{0360-0300}
\urldef\tempurl%
\url{https://doi.org/10.1145/356698.356702}
\showDOI{\tempurl}


\bibitem[\protect\citeauthoryear{Piech, Sahami, Koller, Cooper, and
  Blikstein}{Piech et~al\mbox{.}}{2012}]%
        {piech2012modeling}
\bibfield{author}{\bibinfo{person}{Chris Piech}, \bibinfo{person}{Mehran
  Sahami}, \bibinfo{person}{Daphne Koller}, \bibinfo{person}{Steve Cooper},
  {and} \bibinfo{person}{Paulo Blikstein}.} \bibinfo{year}{2012}\natexlab{}.
\newblock \showarticletitle{Modeling How Students Learn to Program}. In
  \bibinfo{booktitle}{\emph{Proceedings of the 43rd ACM Technical Symposium on
  Computer Science Education}} \emph{(\bibinfo{series}{SIGCSE '12})}.
  \bibinfo{publisher}{Association for Computing Machinery},
  \bibinfo{address}{New York, NY, USA}, \bibinfo{pages}{153--160}.
\newblock
\showISBNx{9781450310987}
\urldef\tempurl%
\url{https://doi.org/10.1145/2157136.2157182}
\showDOI{\tempurl}


\bibitem[\protect\citeauthoryear{Ramey and Fox}{Ramey and Fox}{2020}]%
        {FSFBash2019}
\bibfield{author}{\bibinfo{person}{Chet Ramey} {and} \bibinfo{person}{Brian
  Fox}.} \bibinfo{year}{2020}\natexlab{}.
\newblock \bibinfo{booktitle}{\emph{{The GNU Bash Reference Manual, for Bash,
  Version 5.0}}}.
\newblock Free Software Foundation.
\newblock
\urldef\tempurl%
\url{https://www.gnu.org/savannah-checkouts/gnu/bash/manual}
\showURL{%
\tempurl}


\bibitem[\protect\citeauthoryear{Romero, Ventura, Pechenizkiy, and
  Baker}{Romero et~al\mbox{.}}{2010}]%
        {handbook-edm2010}
\bibfield{editor}{\bibinfo{person}{Cristobal Romero},
  \bibinfo{person}{Sebastian Ventura}, \bibinfo{person}{Mykola Pechenizkiy},
  {and} \bibinfo{person}{Ryan~S.J.d. Baker}} (Eds.).
  \bibinfo{year}{2010}\natexlab{}.
\newblock \bibinfo{booktitle}{\emph{Handbook of educational data mining}}.
\newblock \bibinfo{publisher}{CRC Press}, \bibinfo{address}{Boca Raton, FL,
  USA}.
\newblock
\showISBNx{978-1-4398-0458-2}
\urldef\tempurl%
\url{https://doi.org/10.1201/b10274}
\showDOI{\tempurl}


\bibitem[\protect\citeauthoryear{Security}{Security}{2021}]%
        {metasploit}
\bibfield{author}{\bibinfo{person}{Offensive Security}.}
  \bibinfo{year}{2021}\natexlab{}.
\newblock \bibinfo{booktitle}{\emph{{Metasploit Unleashed}}}.
\newblock OffSec Services Limited.
\newblock
\urldef\tempurl%
\url{https://www.offensive-security.com/metasploit-unleashed/}
\showURL{%
Retrieved November 19, 2021 from \tempurl}


\bibitem[\protect\citeauthoryear{Tenenberg}{Tenenberg}{2019}]%
        {handbook-CER7}
\bibfield{author}{\bibinfo{person}{Josh Tenenberg}.}
  \bibinfo{year}{2019}\natexlab{}.
\newblock \showarticletitle{Qualitative Methods for Computing Education}.
\newblock In \bibinfo{booktitle}{\emph{The Cambridge Handbook of Computing
  Education Research}}, \bibfield{editor}{\bibinfo{person}{Sally~A Fincher}
  {and} \bibinfo{person}{Anthony~V Robins}} (Eds.).
  \bibinfo{publisher}{Cambridge University Press}, \bibinfo{address}{Cambridge,
  United Kingdom}, Chapter~7, \bibinfo{pages}{173--207}.
\newblock
\showISBNx{978-1-108-72189-9}
\urldef\tempurl%
\url{https://doi.org/10.1017/9781108654555}
\showDOI{\tempurl}


\bibitem[\protect\citeauthoryear{Trčka, Pechenizkiy, and van~der Aalst}{Trčka
  et~al\mbox{.}}{2010}]%
        {handbook-edm9}
\bibfield{author}{\bibinfo{person}{Nikola Trčka}, \bibinfo{person}{Mykola
  Pechenizkiy}, {and} \bibinfo{person}{Wil van~der Aalst}.}
  \bibinfo{year}{2010}\natexlab{}.
\newblock \showarticletitle{Process Mining from Educational Data}.
\newblock In \bibinfo{booktitle}{\emph{Handbook of educational data mining}},
  \bibfield{editor}{\bibinfo{person}{Cristobal Romero},
  \bibinfo{person}{Sebastian Ventura}, \bibinfo{person}{Mykola Pechenizkiy},
  {and} \bibinfo{person}{Ryan~S.J.d. Baker}} (Eds.). \bibinfo{publisher}{CRC
  Press}, \bibinfo{address}{Boca Raton, FL, USA}, Chapter~9,
  \bibinfo{pages}{123--142}.
\newblock
\showISBNx{978-1-4398-0458-2}
\urldef\tempurl%
\url{https://doi.org/10.1201/b10274}
\showDOI{\tempurl}


\bibitem[\protect\citeauthoryear{\v{S}v\'{a}bensk\'{y}, Vykopal, Cermak, and
  La\v{s}tovi\v{c}ka}{\v{S}v\'{a}bensk\'{y} et~al\mbox{.}}{2018}]%
        {svabensky2018kypolab}
\bibfield{author}{\bibinfo{person}{Valdemar \v{S}v\'{a}bensk\'{y}},
  \bibinfo{person}{Jan Vykopal}, \bibinfo{person}{Milan Cermak}, {and}
  \bibinfo{person}{Martin La\v{s}tovi\v{c}ka}.}
  \bibinfo{year}{2018}\natexlab{}.
\newblock \showarticletitle{{Enhancing Cybersecurity Skills by Creating Serious
  Games}}. In \bibinfo{booktitle}{\emph{Proceedings of the 23rd Annual ACM
  Conference on Innovation and Technology in Computer Science Education}}
  \emph{(\bibinfo{series}{ITiCSE 2018})}. \bibinfo{publisher}{ACM},
  \bibinfo{address}{New York, NY, USA}, \bibinfo{pages}{194--199}.
\newblock
\showISBNx{978-1-4503-5707-4}
\urldef\tempurl%
\url{https://doi.org/10.1145/3197091.3197123}
\showDOI{\tempurl}


\bibitem[\protect\citeauthoryear{\v{S}v\'{a}bensk\'{y}, Vykopal, Seda, and
  \v{C}eleda}{\v{S}v\'{a}bensk\'{y} et~al\mbox{.}}{2021a}]%
        {Svabensky2021dataset}
\bibfield{author}{\bibinfo{person}{Valdemar \v{S}v\'{a}bensk\'{y}},
  \bibinfo{person}{Jan Vykopal}, \bibinfo{person}{Pavel Seda}, {and}
  \bibinfo{person}{Pavel \v{C}eleda}.} \bibinfo{year}{2021}\natexlab{a}.
\newblock \showarticletitle{Dataset of shell commands used by participants of
  hands-on cybersecurity training}.
\newblock \bibinfo{journal}{\emph{{Data in Brief}}}  \bibinfo{volume}{38}
  (\bibinfo{year}{2021}), 9.
\newblock
\showISSN{2352-3409}
\urldef\tempurl%
\url{https://doi.org/10.1016/j.dib.2021.107398}
\showDOI{\tempurl}


\bibitem[\protect\citeauthoryear{\v{S}v\'{a}bensk\'{y}, Weiss, Cook, Vykopal,
  \v{C}eleda, Mache, Chudovsk\'{y}, and Chattopadhyay}{\v{S}v\'{a}bensk\'{y}
  et~al\mbox{.}}{2021b}]%
        {dataset}
\bibfield{author}{\bibinfo{person}{Valdemar \v{S}v\'{a}bensk\'{y}},
  \bibinfo{person}{Richard Weiss}, \bibinfo{person}{Jack Cook},
  \bibinfo{person}{Jan Vykopal}, \bibinfo{person}{Pavel \v{C}eleda},
  \bibinfo{person}{Jens Mache}, \bibinfo{person}{Radoslav Chudovsk\'{y}}, {and}
  \bibinfo{person}{Ankur Chattopadhyay}.} \bibinfo{year}{2021}\natexlab{b}.
\newblock \bibinfo{booktitle}{\emph{{Dataset: Evaluating Two Approaches to
  Assessing Student Progress in Cybersecurity Exercises}}}.
\newblock Zenodo.
\newblock
\urldef\tempurl%
\url{https://doi.org/10.5281/zenodo.5752288}
\showDOI{\tempurl}


\bibitem[\protect\citeauthoryear{Vykopal, Čeleda, Šeda, Švábenský, and
  Tovarňák}{Vykopal et~al\mbox{.}}{2021}]%
        {my-2021-FIE-KYPO-CSC}
\bibfield{author}{\bibinfo{person}{Jan Vykopal}, \bibinfo{person}{Pavel
  Čeleda}, \bibinfo{person}{Pavel Šeda}, \bibinfo{person}{Valdemar
  Švábenský}, {and} \bibinfo{person}{Daniel Tovarňák}.}
  \bibinfo{year}{2021}\natexlab{}.
\newblock \showarticletitle{{Scalable Learning Environments for Teaching
  Cybersecurity Hands-on}}. In \bibinfo{booktitle}{\emph{Proceedings of the
  51st IEEE Frontiers in Education Conference}} \emph{(\bibinfo{series}{FIE
  '21})}. \bibinfo{publisher}{IEEE}, \bibinfo{address}{New York, NY, USA},
  \bibinfo{pages}{1--9}.
\newblock
\urldef\tempurl%
\url{https://www.muni.cz/en/research/publications/1783808}
\showURL{%
\tempurl}


\bibitem[\protect\citeauthoryear{Weiss, Boesen, Sullivan, Locasto, Mache, and
  Nilsen}{Weiss et~al\mbox{.}}{2015}]%
        {edurange_SIGCSE2015}
\bibfield{author}{\bibinfo{person}{Richard Weiss}, \bibinfo{person}{Stefan
  Boesen}, \bibinfo{person}{James~F. Sullivan}, \bibinfo{person}{Michael~E.
  Locasto}, \bibinfo{person}{Jens Mache}, {and} \bibinfo{person}{Erik Nilsen}.}
  \bibinfo{year}{2015}\natexlab{}.
\newblock \showarticletitle{Teaching Cybersecurity Analysis Skills in the
  Cloud}. In \bibinfo{booktitle}{\emph{Proceedings of the 46th ACM Technical
  Symposium on Computer Science Education}} \emph{(\bibinfo{series}{SIGCSE
  '15})}. \bibinfo{publisher}{Association for Computing Machinery},
  \bibinfo{address}{New York, NY, USA}, \bibinfo{pages}{332--337}.
\newblock
\showISBNx{9781450329668}
\urldef\tempurl%
\url{https://doi.org/10.1145/2676723.2677290}
\showDOI{\tempurl}


\bibitem[\protect\citeauthoryear{Weiss, Locasto, and Mache}{Weiss
  et~al\mbox{.}}{2016}]%
        {weiss2016}
\bibfield{author}{\bibinfo{person}{Richard Weiss}, \bibinfo{person}{Michael~E.
  Locasto}, {and} \bibinfo{person}{Jens Mache}.}
  \bibinfo{year}{2016}\natexlab{}.
\newblock \showarticletitle{{A Reflective Approach to Assessing Student
  Performance in Cybersecurity Exercises}}. In
  \bibinfo{booktitle}{\emph{Proceedings of the 47th ACM Technical Symposium on
  Computing Science Education}} \emph{(\bibinfo{series}{SIGCSE '16})}.
  \bibinfo{publisher}{ACM}, \bibinfo{address}{New York, NY, USA},
  \bibinfo{pages}{597--602}.
\newblock
\showISBNx{978-1-4503-3685-7}
\urldef\tempurl%
\url{https://doi.org/10.1145/2839509.2844646}
\showDOI{\tempurl}


\bibitem[\protect\citeauthoryear{Weiss, Turbak, Mache, and Locasto}{Weiss
  et~al\mbox{.}}{2017}]%
        {weiss2017magazine}
\bibfield{author}{\bibinfo{person}{Richard Weiss}, \bibinfo{person}{Franklyn
  Turbak}, \bibinfo{person}{Jens Mache}, {and} \bibinfo{person}{Michael~E
  Locasto}.} \bibinfo{year}{2017}\natexlab{}.
\newblock \showarticletitle{{Cybersecurity Education and Assessment in
  EDURange}}.
\newblock \bibinfo{journal}{\emph{IEEE Security \& Privacy}}
  \bibinfo{volume}{15}, \bibinfo{number}{3} (\bibinfo{year}{2017}),
  \bibinfo{pages}{90--95}.
\newblock
\showISSN{1558-4046}
\urldef\tempurl%
\url{https://doi.org/10.1109/MSP.2017.54}
\showDOI{\tempurl}


\end{thebibliography}

\end{document}